\documentclass[aps,pre,superscriptaddress,twocolumn,amsmath,amssymb,showpacs]{revtex4}
\usepackage{graphicx}
\usepackage{dcolumn}
\usepackage{bm}
\newcommand{\ignore}[1]{}
\usepackage{epsfig}
\usepackage{color}
\usepackage{amsmath,amssymb}

\begin{document}

\title{Optimal recruitment strategies for groups of interacting walkers with leaders}

\author{Ricardo Mart\'inez-Garc\'ia}
\affiliation{IFISC, Instituto de F\'isica Interdisciplinar y Sistemas Complejos (CSIC-UIB), E-07122 Palma de Mallorca, Spain}
\affiliation{Department of Ecology and Evolutionary Biology, Princeton University, Princeton NJ 08544-1003, USA}
\author{Crist\'obal L\'opez}  
\affiliation{IFISC, Instituto de F\'isica Interdisciplinar y Sistemas Complejos (CSIC-UIB), E-07122
  Palma de Mallorca, Spain}
\author{Federico Vazquez}
\affiliation{Instituto de F\'isica de L\'iquidos y Sistemas Biol\'ogicos UNLP-CONICET, 1900 La Plata, Argentina}

\begin{abstract}
We introduce a model of interacting random walkers on a finite
one dimensional chain with absorbing boundaries or targets at the
ends.  Walkers are of  two types: \emph{informed} particles that move
ballistically towards a given target, and diffusing 
\emph{uninformed} particles that are biased towards close informed
individuals. This model mimics the dynamics of hierarchical groups
  of animals, where an informed individual tries to persuade and lead
  the movement of its conspecifics.
We characterize the success of this persuasion by the first-passage probability of the uninformed
particle to the target, and we interpret the speed of the informed
particle as a strategic parameter that the particle can tune to
maximize its success.  We find that the success probability is non-monotonic,
reaching its maximum at an intermediate speed whose value increases with the
diffusing rate of the uninformed particle.  When two different groups of
informed leaders traveling in opposite directions compete, usually
the largest group is the most successful.  However, the minority can
reverse this situation and become the most probable winner by
following two different strategies: increasing its attraction strength
or adjusting its speed to an optimal value relative to the majority's speed.  
\end{abstract}


\maketitle

\section{Introduction}
\label{secintro}

Animal species may show democratic behavior when decisions within a group are shared
by most or all of its members  \cite{Conradt2005,Couzin2011,Conradt2012}, as it
has been reported in honey bees \cite{seeley1999group} or fish schools \cite{Couzin2011}. 
On the contrary, unshared or despotic decisions appear in hierarchical species, where
one or a few leaders dominate and
 influence its conspecifics  \cite{chase1974models}. 
This social structure has been widely reported in several animal species such as pigeons  \cite{Nagy2010},
 cattle \cite{Reinhardt1983}, dwarf mongooses \cite{Rasa1983},
dolphins \cite{Smolker2000} and many primates as for instance rhesus macaques \cite{Reinhardt1987},
black spider monkeys \cite{M.G.M.vanRoosmalen1980}, 
chimpanzees \cite{Boesch1989} or yellow baboons \cite{Rhine1975} (see \cite{gauthreaux1978ecological} and references
therein). 

In the context of animal movement, which is the 
focus of this paper, leaders usually initiate 
the movement,  persuading the rest of the group to follow their
steps to maintain cohesion.  Thus, communication together with
individual experience induces the formation of groups
\cite{Miller2013,Eftimie2007,couzin2003self,krause2002living}.  A
leader decides where to move based on its own knowledge and,
due to its influence on the group, it may recruit 
some conspecifics who follow its movements. This displacement strategy
may be used to increase  defense, predation or foraging success
\cite{Sueur2008}.

In this paper we  focus on the situation where a leader 
has beforehand information about the location of resources
and moves towards them. This informed individual, 
characterized by a persuading force whose strength might depend on
some of its attributes like size, personality, etc.,
is able to tune its speed in order to increase its influence on the movement of the group and recruit more conspecifics. 
Thus, if for any reason animals do not follow the leader, the group
may split into smaller groups, as it
happens in the so-called fission-fusion societies observed in
elephants, dolphins, some ungulates and primates
\cite{kerth2010group}.
However, following the informed leader may
incidentally benefit 
uninformed individuals that, apart from being protected by the group, may obtain other advantages like knowledge about
the location of resources.

 We quantify these benefits in terms of the first-passage probability 
of arriving at the target and explore how the 
interaction with a leader may affect searching processes.
We analyze the likelihood that an uninformed
member finds a specific target, and how that depends on the speeds and
interaction parameters of the informed individuals. In a second part,
we  extend our scenario to the case where several leaders compete for
an uninformed individual and address the question of when a minority group of informed individuals
can beat a majority \cite{Couzin2011,IainD.Couzin2005,Conradt2005}.

This scenario  differs from previous studies
where a whole population searches collectively for food.
This last behavior is more frequent in democratic species as it has been reported
in recent works where fish schools explore complex environments \cite{Berdahl2013}.
Conspecifics act as an ``array of sensors'' to pool their information 
and better average their movement decisions \cite{Torney2009, Hoare2004, Grunbaum1998}.

From a physical point of view, first-passage  problems are of
fundamental relevance for stochastic  processes,  and have
  been studied in systems of non-interacting Brownian particles in  
various scenarios, such as those subject to external potentials,
with different boundary conditions or under different diffusion coefficients \cite{Redner,Schehr2012,BenNaim2010,Krapivsky-2012}.
Some recent related work has dealt with interacting active particles but focusing on the
collective properties of the system \cite{Peruani} and pattern formation \cite{Eftimie2007}, rather than in searching (first-passage properties) tasks.
An interaction mechanism and its influence on high-order statistics, the general aim of this paper,
has therefore not been thoroughly studied yet.
Only few recent studies have investigated its effect on first-passage properties
within a searching context \cite{Martinez-Garcia2013b,Tani2014},
concluding that the optimal situation where the group is benefited as a
whole is a mixture of independent searching and joining other members
in the search.  It was shown that this conclusion is independent on
the mobility pattern, either L\'evy or Brownian
\cite{Martinez-Garcia2014a}. 
 
The paper is organized as follows. 
In Section \ref{model} we present the general characteristics of the model. 
In Section \ref{oneinformed} we
analyze the simplest case of one informed and one uninformed interacting
individuals. In Section \ref{twoinformed} we study the case of two 
informed leaders competing for recruiting one uninformed partner. Section 
\ref{competegroups} explores the effect of having more than one leader in one of the groups.
The paper ends with a summary and conclusions in Section \ref{sec:conclusions}.

\section{General considerations of the model}
\label{model}

 In its simplest version, the model  mimics the  behavior of two
interacting animals, a leader and its follower.  The leader or
\emph{informed} animal, called particle $I$, 
moves with a constant speed towards the location of food resources.  Because of
its leadership, $I$ ``drags'' its follower or \emph{uninformed}
partner (particle $U$) in the direction of the food target.  This ``persuasive''
interaction only occurs 
when $U$ and $I$ are close enough, so they can communicate, increasing
the chances that $U$ finds the resources.

In mathematical terms, particles $U$ and $I$ perform a random walk on a
discrete one-dimensional space represented by a chain with sites
labeled from $n=0$ to $n=N$.  Both particles are initially placed at the center
site $N/2$, whereas the target is located at the extreme site $N$.
Particle $I$ jumps with rate $k_i$ to its right neighboring site (left jumps are
forbidden). Note that since the dynamics is defined on continuous
time, a jumping rate $k_i$ implies that the probability of jumping
in an infinitesimal time step $dt$ is $k_i dt$.  Particle $U$
jumps with a rate $k_u$ that depends on whether its distance $d=|x_u-x_i|$ to the
particle $I$ is smaller or larger than an interaction range $R>0$.
We denote here by $x_u$ and $x_i$ the positions of $U$ and 
$I$ respectively.  When $d>R$ particles do not interact,
thus $U$ jumps right or left with equal rates $k_u^+=1$ and $k_u^-=1$,
respectively. However, when $d \le R$ particle $I$ attracts $U$, which jumps
towards $I$ with rate $k_u=1+k_0$ and away from $I$ with rate
$k_u=1$. The parameter $k_0 > 0$ measures the strength of the recruitment attraction.
Formally, right and left $U$'s jumping rates are (see Fig.~\ref{jumping-1}):

\begin{figure}
\centering 
\includegraphics[width=0.47\textwidth]{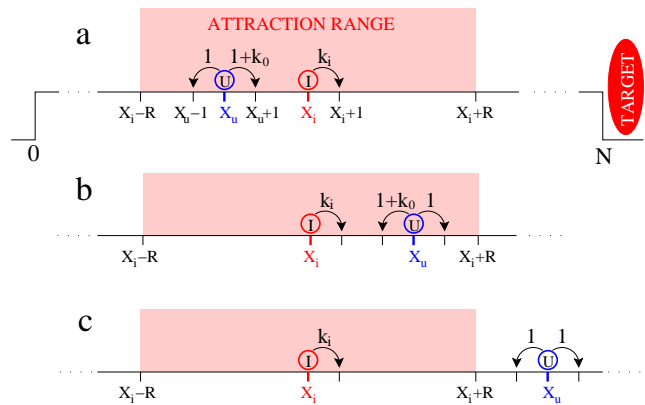}
\caption{(Color online). Illustration of the model's dynamics.
  Informed particle $I$ jumps rightwards with rate $k_i$ until it hits
  the food target at site $N$.  (a) and (b) When the
  uninformed particle $U$ is inside $I$'s attraction range $R$ (red
  shaded region), it jumps with rate $k_u=1+k_0$ towards $I$ and with rate
  $k_u=1$ away from  $I$.  (c) When it is outside that range, $U$ jumps right and left with rate $k_u^\pm=1$.
  $U$ stops when it reaches either the absorbing site
  $0$ or $N$.}  
\label{jumping-1}
\end{figure}

\begin{eqnarray*}
 k_u^+ = \left\{ \begin{array}{lll}
1 & \mbox{for} & 0 < x_u < x_i-R, \\
1+k_0 & \mbox{for} & x_i-R \le x_u < x_i, \\
1 & \mbox{for} & x_i \le x_u < N.
\end{array}
\right.
\end{eqnarray*}

\begin{eqnarray*}
 k_u^- = \left\{ \begin{array}{lll}
1 & \mbox{for} & 0  < x_u \le x_i, \\
1+k_0 & \mbox{for} & x_i < x_u \le x_i+R, \\
1 & \mbox{for} & x_i+R < x_u < N.
\end{array}
\right.
\end{eqnarray*}

In other words, particle $I$ moves ballistically towards a known target 
located at site $N$, while particle $U$ experiments a bias $k_0$
towards $I$ when it is within $I$'s interaction range $R$, and moves
as a symmetric random walker as long as it is outside this range.
Finally, when both particles occupy the same site, $U$ jumps with equal 
rates $k_u^+=k_u^-=1$.  Particle $U$
stops walking when it reaches either site $0$ or 
$N$ (absorbing sites of the system), and particle $I$ stops at site
$N+R$, so that the communication between $U$ and 
$I$ is switched off once $I$ finds the target at $n=N$.  Note
that if the motion of $U$ were independent
of $I$ and performing a symmetric random walk, $U$ would have the
same likelihood to reach either end of the chain.  However, given that
$I$ moves to the right and ``attracts'' $U$ when they are close enough,
one can view the dynamics as $U$ receiving ``effective kicks''
to the right and, therefore, one expects $U$ to have a preference for
the target located at $n=N$.

As indicated in the introduction,  this model aims to tackle some of the fundamental  
questions about the relationship between animal interactions and
searching processes.  What is the effect of leadership
interactions on foraging success? How does the probability of
reaching the food target depend on the speed and diffusion of
both leading and recruited particles?  What happens in a more complex and realistic
scenario with many competing leaders?  To
address these questions we study in the next three sections the cases of
one uninformed particle and one or several informed particles.  We focus
on the probability that $U$ reaches the right target $N$, or
first-passage probability $F^{\mbox{\tiny N}}$ to target $N$.  
Our aim is to explore how this quantity depends on the speed $k_i$
of $I$, its attraction range $R$ and the bias $k_0$.  Within a
biological context, $k_i$ 
can be seen as a strategic parameter that an informed animal wants to
tune in order to optimize its recruitment strength and maintain group cohesion.

\section{One informed and one uninformed particles}
\label{oneinformed}

\begin{figure}
\centering  \includegraphics[width=0.4\textwidth]{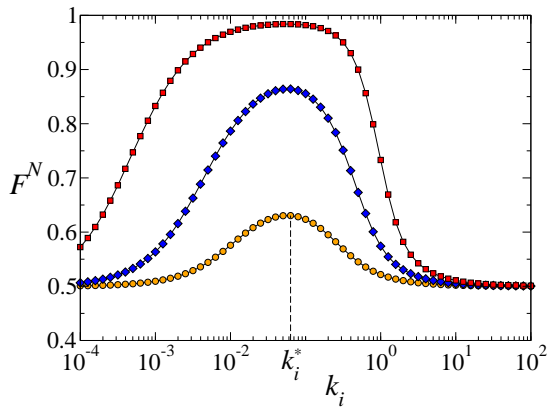}
\caption{(Color online).  Probability $F^{\mbox{\tiny N}}$ that the
  uninformed particle $U$ reaches target $N$ vs the jumping rate
  $k_i$ of the informed particle, for internal strengths 
  $k_{0}=0.2$ $(\bullet)$, $k_{0}=0.5$ $(\blacklozenge)$ and $k_{0}=1.0$
  $(\blacksquare)$.  $F^{\mbox{\tiny N}}$ is maximum at $k_i^*$.  Simulations
  correspond to a chain of length $N=100$ and an attraction range of
  the informed particle $R=10$.}
\label{full-1}
\end{figure}

We start performing numerical
simulations of the dynamics described in Section \ref{model}.  As we
see in Fig.~\ref{full-1}, the first-passage probability (FPP)
$F^{\mbox{\tiny N}}$ is non-monotonic in the jumping rate or speed
$k_i$ of particle $I$, showing its maximum at intermediate speeds.  That
is, there is an \emph{optimal speed} denoted by $k_i^*$, for which
the success probability $F^{\mbox{\tiny N}}$ of $U$ is maximum.

\begin{figure}
\centering  \includegraphics[width=0.4\textwidth]{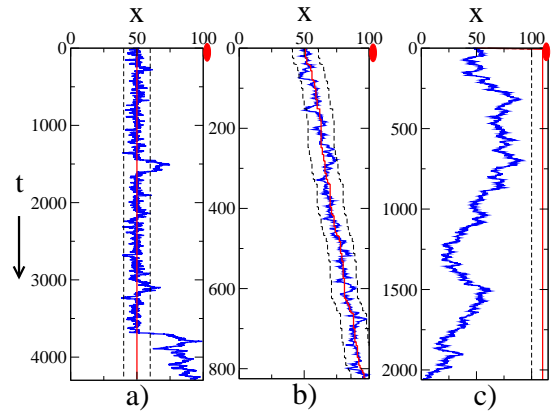}
\caption{(Color online) Individual realizations of the searching
  process representing typical trajectories of the particles
  in three different regimes of the model: (a) small $k_i$, (b)
  optimal search $k_i=k_i^*=0.06$ and (c) large $k_i$. Space and time
  are displayed on the $x$ and $y$ axis, respectively, and the target
  at site $N=100$ is denoted by a red ellipse.  Parameters:
  $N=100$, $R=10$ and $k_0=0.2$.  In all plots black dashed lines represent the
  interaction range, which is centered at the red solid line representing
  the trajectory of the informed particle, whereas the remaining blue solid
  line shows the trajectory of the uninformed particle.}
\label{trajectories}
\end{figure}

We can distinguish three different regimes depending on $I$'s speed:
(i) the low speed regime, (ii) the optimal
regime $k_i \simeq k^{*}_{i}$, and (iii) the high speed regime.
Typical trajectories of each case are shown
in Figs.~\ref{trajectories}(a), (b) and (c), respectively. 
In regime (i) $I$ remains almost static [see Fig.~\ref{trajectories} (a)],
so jumping rates of $U$ are symmetric around the center of the
chain, and $U$ reaches target $N$ with probability
$F^{\mbox{\tiny N}} \simeq 1/2$.  In 
this case, $I$ interacts many times with $U$, but its overall effect is
null because of the symmetry of its position and interaction.  
In the other extreme, regime (iii), the interaction between $U$ and
$I$ is negligible, given that $I$ moves very fast and quickly leaves 
$U$ out of its interaction range [see Fig.~\ref{trajectories}(c)]. Therefore,
$U$ performs a symmetric random walk leading to 
$F^{\mbox{\tiny N}} \simeq 1/2$.  Finally, in the 
intermediate regime (ii), $I$ moves at a speed that  ``traps'' $U$
inside the interaction range most of its way to the target   [see 
Fig.~\ref{trajectories}(b)].  This ``right speed'' is not too
fast to overtake and leave $U$ behind, but also not too slow to have no
effective drag on $U$.

\begin{figure}
\centering 
\includegraphics[width=0.40\textwidth]{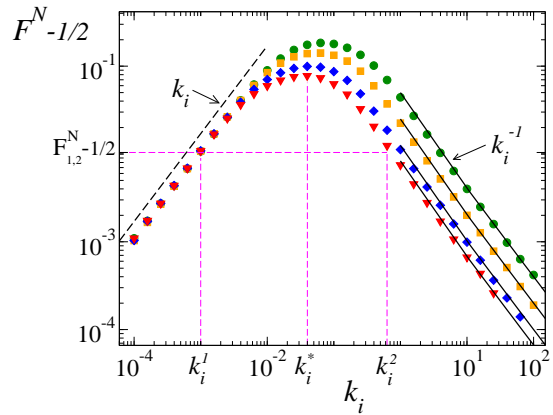}
\caption{(Color online) First-passage probability of the uninformed
  particle to target $N$, $F^{\mbox{\tiny N}}$, vs the speed of the
  informed particle $k_i$ on a log-log plot, for chains of
  length $N=50$ $(\bullet)$, 
  $N=100$ $(\blacksquare)$, $N=200$ $(\blacklozenge)$ and $N=300$
  $(\blacktriangledown)$, with 
  $k_0=0.2$ and $R=10$.  $F^{\mbox{\tiny N}}$ is shifted by $1/2$ to
  show the asymptotic behaviors $F^{\mbox{\tiny N}}-1/2 \sim k_i$ and
  $F^{\mbox{\tiny N}}-1/2 \sim k_i^{-1}$ in the small and large $k_i$
  limits, respectively.  Solid lines represent the high speed approximation
  from Eq.~(\ref{FN-high-ki}), while the dashed line 
  corresponds to the low speed approximation (\ref{FN-low-ki}).
  Vertical dashed lines denote the positions of the maximum $k_i^*$
  and of two arbitrary speeds $k_i^1$ and $k_i^2$ with the same
  probability $F_{1,2}^{\mbox{\tiny N}}$.} 
\label{prob-ki}
\end{figure}

Figure~\ref{prob-ki} shows that $F^{\mbox{\tiny N}}$ increases from $1/2$
linearly with $k_i$ in the very low speed limit, while it
decays to $1/2$ as $k_i^{-1}$ in the very large $k_i$ limit.  In the next
three subsections we explore each regime in more detail and provide
analytic estimations. 

\subsection{The large $k_i$ regime}

In the very high speed limit $k_i \to \infty$, the informed particle
$I$ reaches target $N$ in a very short time.  Thus, the uninformed
particle $U$ does not have the chance to make even a single step,
staying at the center site of the chain, which we will call site $c \equiv N/2$ from
now on. More precisely, given that $I$ jumps in a mean time $1/k_i$, 
when $k_i \gg (2+k_0)(N/2+R)$, $I$ arrives to the target in a mean time 
$(N/2+R)/k_i$, which is much smaller than the typical time
$(2+k_0)^{-1}$ that $U$ needs to make a single jump.  This
extreme behavior is shown in  Fig.~\ref{trajectories}(c).  Hence,
after $I$ reaches the target, $U$ performs a symmetric random walk
starting from $n=c$, and hitting site $N$ with probability $F^{\mbox{\tiny
    N}}=1/2$.  Now, following the same reasoning, we consider the 
$k_i \gg 2+k_0$ limit, that is, $k_i$ lower than the extreme
value considered before but still high.  This corresponds to the case
in which the bias in $I$ is much larger than the diffusion of $U$.
Given that $I$ moves right much faster than $U$, $U$ makes a few
steps before leaving the box through the left side.  Therefore, we
can assume that 
once $U$ leaves the box it never comes back in again, because it is
very unlikely that $U$ diffusing with a very low rate can catch the very
fast moving box.  After $U$ leaves the box at a given position $x$, it
starts a pure diffusion motion, reaching target $N$ with a probability
that increases linearly as we approach to site $n=N$ \cite{Redner,gardiner}
\begin{equation}
F^{\mbox{\tiny N}}=\frac{x}{N}.
\label{FxN}
\end{equation}
In the mean time $\Delta t$ that $U$ takes to leave the box, the box
travels a mean distance $k_i \, \Delta t$ and, therefore, $U$ leaves
the box at position  
\begin{equation}
x=\frac{N}{2}+k_i \, \Delta t - R - 1,
\label{separation}
\end{equation}
where $R+1$ accounts for the distance between $U$ and the center of
the box at the exit moment.  The exit time $\Delta t$ can be
calculated working in the reference frame of the box, where the relative
position of $U$ respect to the box's center is $x'_u = x_u - x_i$, thus 
$-R \le x'_u \le R$ inside the box.  Then $U$ jumps in the box's reference 
frame with rates 
\begin{eqnarray}
k_u'^+(n) &=& \left\{ \begin{array}{ll}
1+k_0 & \mbox{for $-R \le n \le -1$,} \\
1 & \mbox{for $0 \le n \le R$.} \end{array} \right. \nonumber \\
k_u'^-(n) &=& \left\{ \begin{array}{ll}
1+k_i & \mbox{for $-R \le n \le 0$,} \\
1+k_0+k_i & \mbox{for $1 \le n \le R$.}\end{array} \right. 
\label{kn+-box}
\end{eqnarray}
In the limit $k_i \gg 2+k_0$ considered here, $U$ moves 
ballistically to the left under a strong bias $k_i-k_0$, when it is
seen from the perspective of the box. Thus the mean
exit time corresponds to that of a ballistic motion with
$R+1$ left steps: $1$ step and then $R$ steps with effective left
rates $k_i$ and $k_i-k_0$, respectively,  
\begin{equation}
\Delta t \simeq \frac{1}{k_i} + \frac{R}{k_i-k_{0}}.
\label{time}
\end{equation}
Plugging expression (\ref{time}) for $\Delta t$ into
Eq.~(\ref{separation}) gives 
\begin{equation}
x \simeq \frac{N}{2}+\frac{k_{0}R}{k_i-k_{0}},
\label{x}
\end{equation}
and using this expression for $x$ in Eq.~(\ref{FxN}), we finally obtain 
\begin{equation}
F^{\mbox{\tiny N}}=\frac{1}{2}+\frac{k_{0}R}{N(k_i-k_{0})}.
\label{FN-high-ki}
\end{equation}
As we show in Fig.~\ref{prob-ki}, Eq.~(\ref{FN-high-ki}) reproduces
very well the behavior of $F^{\mbox{\tiny N}}$ from numerical
simulations in the $k_i \gg 2+k_0 = 2.2$ regime.  Discrepancies
between theory and simulations start to be important for $k_i \lesssim
1$.  In summary, we showed that the FPP approaches to $1/2$ as $k_i^{-1}$ in the
large $k_i$ limit.

\subsection{The small $k_i$ regime}

In the simplest case $k_i=0$, $I$ remains fixed at the center site
$c=N/2$.  The interval $[c-R,c+R]$, where $U$ feels the presence of
$I$, defines an ``attracting box'' of length $2R$ centered at $I$'s
position $x_i=c$.  One can see the dynamics as $U$ performing a random walk on a
chain with quenched right $k_u^+(n)$ and left $k_u^-(n)$ site-dependent
jumping rates (see Fig. \ref{jumping-1})
\begin{eqnarray}
k_u^+(n) = \left\{ \begin{array}{ll}
1+k_0 & \mbox{for $c-R \le n \le c-1$,} \\
1 & \mbox{otherwise.}\end{array} \right. \nonumber \\
k_u^-(n) = \left\{ \begin{array}{ll}
1+k_0 & \mbox{for $c+1 \le n \le c+R$,} \\
1 & \mbox{otherwise.}\end{array} \right.
\label{kn+-}
\end{eqnarray}
Because of the symmetry of rates around $c$ and of
the initial condition $x_u(0)=x_i(0)=c$, $U$ has the same chance to
hit both targets, 
thus $F^{\mbox{\tiny N}}=1/2$.  Now, if $k_i$ is larger than zero, $U$ rates
$k_u^{\pm}(n)$ change every time the box makes a step to the right.
Therefore, in this situation rates vary not only along the chain but
also on time. This case is hard to analyze, so we focus here on the
simplest non-trivial limit of very low $k_i$.  

We consider the situation in which the typical time $1/k_i$ that $I$
takes to make a single step is much smaller than the mean time $T_c$
that $U$ needs 
to reach an end of the chain, starting from site $c$.  An exact expression
for $T_c$ is obtained in Appendix \ref{appen-T} [see also 
  Eq.~(\ref{T})].  During the time $T_c$, $I$ makes $n$ 
steps with probability $p_n = \frac{k_i^n T_c^n}{n!} e^{-k_i T_c}$.  Then,
in the $k_i \ll 1/T_c$ limit these probabilities are reduced to 
$p_0 = 1-k_i T_c + \mathcal O(k_i^2 T_c^2)$, $p_1 = k_i T_c + \mathcal
O(k_i^2 T_c^2)$, $p_n = k_i^n T_c^n + \mathcal O(k_i^{n+1} T_c^{n+1})$ for 
$n \ge 2$.  Therefore, neglecting terms of order $k_i^2 T_c^2 \ll 1$ and
higher, only two events are statistically possible: (1) the box
moves one step to site $n=c+1$, with probability $P_{c+1}=p_1=k_i
T_c$, or (2) the box does not move, staying at the initial center site
$n=c$, with probability $P_c=p_0=1-k_i T_c$.  That is, we sort all
possible realizations
of the dynamics into two 
classes, those in which $I$ jumps once before $U$ exits the chain and those in 
which $U$ exits before $I$ makes any jump.  If we denote by 
$F_{c+1}^{\mbox{\tiny N}}$ ($F_c^{\mbox{\tiny N}}$) the FPP to
target $N$ in the first (second) event, then the FFP can be calculated
as
\begin{equation}
 F^{\mbox{\tiny N}}= P_{c} \, F_{c}^{\mbox{\tiny N}} + P_{c+1} \,
 F_{c+1}^{\mbox{\tiny N}}.  
\label{E}
\end{equation}
In the second event (box does not move), the FPP to target $N$ is
simply $F_c^{\mbox{\tiny N}}=1/2$, as mentioned before.   
Then, Eq.~(\ref{E}) is reduced to
\begin{equation}
 F^{\mbox{\tiny N}}= \frac{1}{2} + \left(  F_{c+1}^{\mbox{\tiny
     N}}-\frac{1}{2} \right) k_i \, T_c. 
\label{E-1}
\end{equation}
To calculate $T_c$ we take advantage of the symmetry of $U$'s jumping rates 
around $c$, and map the chain $[0,N]$ with absorbing boundaries at
sites $0$ and $N$ to a chain $[c,N]$ with reflecting and absorbing
boundaries at sites $c$ and $N$, respectively (see Appendix
\ref{appen-T}).  We obtain   
\begin{equation}
T_c = \tau_e + \frac{(\tilde{N}-2) \left[ (2+k_0)(1+k_0)^R - 2 \right]}{4 k_0}
+ \frac{\tilde{N}(\tilde{N}-2)}{8}, 
\label{T}
\end{equation}
where we have defined $\tilde{N} \equiv N-2R$, and 
\begin{equation}
\tau_e = \frac{(2+k_0) \left[ (1+k_0)^{R+1} -1 \right] - 2 k_0 (R+1)}{2k_0^2}
\label{taue}
\end{equation}
is the mean time that $U$ takes to escape from the box, starting from
the center and with the box fixed (see Appendix \ref{appen-taue}).
The estimation of $F_{c+1}^{\mbox{\tiny N}}$ involves many steps,
which we develop in Appendix \ref{appen-E} for the interested reader.
The approximate final result is   
\begin{eqnarray}
 F_{c+1}^{\mbox{\tiny N}} \simeq \frac{1}{2} + \frac{8\, \tau_e}
 {\tilde{N}(8 \tau_e + \tilde N) }.
\label{Fc1}
\end{eqnarray}
Plugging this expression for $ F_{c+1}^{\mbox{\tiny N}}$ into 
Eq.~(\ref{E-1}) we obtain
\begin{equation}
 F^{\mbox{\tiny N}}\simeq \frac{1}{2} + \frac{8\, \tau_e \, k_i
   \,T_c}{\tilde{N}(8\tau_e+ \tilde{N})}.  
\label{E-2}
\end{equation}
For the parameter values used in simulations $N=50,100,200$ and 
$300$, $R=10,14,18$ and $22$, $k_0=0.2, 0.5$ and $0.8$, we can
simplify Eqs.~(\ref{T}) and (\ref{taue}) for $T_c$ and
$\tau_e$, by retaining only the 
leading terms.  For instance, the factor $(1+k_0)^{R+1}$ becomes dominant in
Eq.~(\ref{taue}), thus we get
\begin{equation}  
\tau_e \simeq \frac{(2+k_0)(1+k_0)^{R+1}}{2k_0^2}.
\label{taue-2}
\end{equation}
Using the simplified Eq.~(\ref{taue-2}), we can rewrite Eq.~(\ref{T})
for $T_c$ as
\begin{equation}
T_c \simeq \tau_e + \frac{k_0 (\tilde{N}-2) \tau_e}{2(1+k_0)} + 
\frac{\tilde N (\tilde N-2)}{8}.
\label{T-2}
\end{equation}
One can also check that $\tau_e \gg \tilde N$ for most combinations
of $N$, $R$ and $k_0$.  Then, plugging Eq.~(\ref{T-2}) for
$T_c$ into Eq.~(\ref{E-2}) and expanding to first order in
$\tilde N / \tau_e$ we get
\begin{equation}
F^{\mbox{\tiny N}}\simeq \frac{1}{2} + \frac{k_0 \tau_e k_i}{2(1+k_0)} 
\left[ 1+\frac{(2+k_0) \tilde{N}}{8 k_0 \tau_e} \right],
\label{E-3}
\end{equation}
where we have used $\tilde N \gg 2(1+k_0)/k_0$.  Finally, replacing
$\tau_e$ from Eq.~(\ref{taue-2}) into Eq.~(\ref{E-3}), and keeping
only the leading term we arrive to  
\begin{equation}
 F^{\mbox{\tiny N}}\simeq \frac{1}{2} + \frac{(2+k_0)(1+k_0)^R \,k_i}{4 k_0}. 
\label{FN-low-ki}
\end{equation}

\begin{figure}
\centering 
\includegraphics[width=0.47\textwidth]{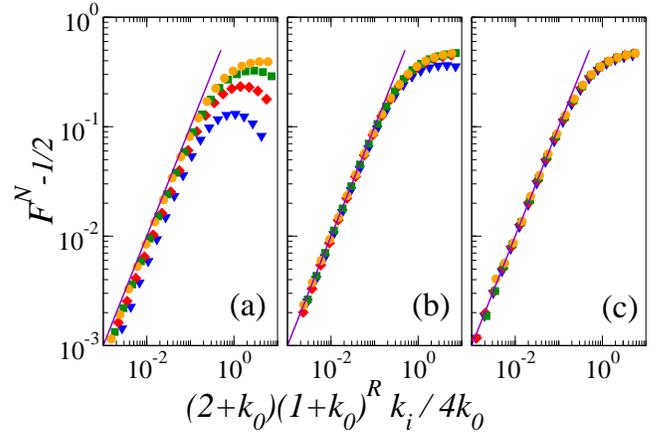}
\caption{(Color online) Scaling of the first-passage probability of
  the uninformed particle $F^{\mbox{\tiny N}}$ in the limit of low
  speed $k_i$ of the informed particle, for a chain of length $N=100$
  and four values of the attraction range: $R=10
  \,(\blacktriangledown)$, $R=14 \, 
  (\blacklozenge)$, $R=18 \,(\blacksquare)$ and $R=22 \,(\bullet)$.
  Each panel corresponds to a different $k_0$: (a) $k_0=0.2$, (b)
  $k_0=0.5$ and (c) $k_0=0.8$.  $y$ and $x$ axis were rescaled in
  order to compare simulation results with the analytic expression  
  (\ref{FN-low-ki}) (straight solid lines), which corresponds to the
  straight line $y=x$ in these rescaled axis.  Agreement between
  simulation and theory improves as $k_0$ and $R$ increase.} 
\label{F-ki-R-k0}
\end{figure}

In Fig.~\ref{F-ki-R-k0} we show the asymptotic behavior of  the FPP in
the small $k_i$ limit.  $F^{\mbox{\tiny N}}$ is shifted by $1/2$ and
the speed $k_i$ is rescaled according to the analytic estimation 
from Eq.~(\ref{FN-low-ki}), denoted by the straight solid line.
Simulations correspond to a chain of length $N=100$, while each of
the four curves is for a different value of the attraction range $R$.
Panels (a), (b) and (c) correspond to strengths $k_0=0.2, 0.5$ and
$0.8$, respectively. We observe that as $R$ and $k_0$ increase, 
the agreement between numerical results and the analytic curve
from Eq.~(\ref{FN-low-ki}) improves.  This is because as $R$ and $k_0$ get
larger, the term $(1+k_0)^{R+1}$ becomes more dominant, and thus the approximate
expression for $\tau_e$ from Eq.~(\ref{taue-2}) and the assumption $\tau_e
\gg \tilde N$ improve. In
Fig.~\ref{prob-ki}, where we plot the FPP vs $k_i$ for various system
sizes $N$, we can see that $F^{\mbox{\tiny N}}$ does
not depend on $N$ for low $k_i$, as predicted by
Eq.~(\ref{FN-low-ki}).  In summary, 
the FPP to target $N$ in the small $k_i$ limit is proportional to
$k_i$, exponential of $R$ and independent of the chain's length $N$.

\subsection{Optimal regime $k_i \simeq k_i^*$ }

We have studied in the last two subsections the limiting cases where
the informed particle $I$ moves either too slow
[Fig.~\ref{trajectories}(a)] or too fast
[Fig.~\ref{trajectories}(c)], to guide the movement of the
uninformed searcher $U$.  In this subsection we investigate the
properties of the optimal searching regime [Fig.~\ref{trajectories}
  (b)], where the probability  that $U$ reaches the right target
is maximum, which happens at intermediate values of $I$'s speed $k_i$.

\begin{figure}
\centering  \includegraphics[width=0.45\textwidth]{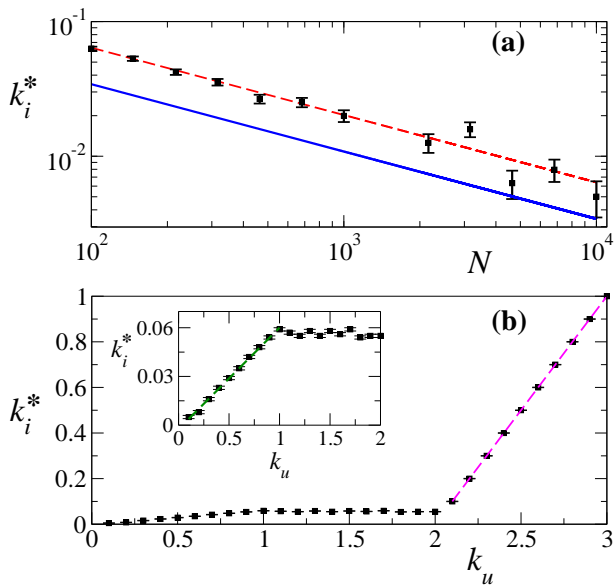}
\caption{(Color online) (a) Speed of the informed particle $k_i^*$ that
  maximizes $F^{\mbox{\tiny N}}$ vs system size $N$ on a double
  logarithmic scale, obtained from numerical simulations, with $R=10$
  and $k_{0}=0.2$. The solid line corresponds to
    Eq.~(\ref{ki-optimal}), while the dashed line shows the scaling 
    $k_i^* \sim N^{-1/2}$.
  (b) $k_i^*$ as a function of the free jumping rate 
  $k_u$ of the uninformed particle on a chain of length $N=100$.  
  The dashed line is 
  $k_i^* = k_u-2$.  Inset: in the low $k_u$ regime, data is well fitted
  by $k_i^* = 0.062\, k_u$ (dashed line).  $k_i^*$ is constant in the
  $1 \le k_u \le 2.062$ interval.} 
\label{scaling-optki}
\end{figure}

A rough estimation of the optimal speed $k_i^*$ can be obtained if we
assume that the asymptotic behaviors of the FPPs in the small and
large $k_i$ limits from Eqs.~(\ref{FN-low-ki}) and 
(\ref{FN-high-ki}), respectively, intersect near the maximum. 
Equating these two expressions and solving for $k_i$ leads to  
\begin{equation}
k_i^*=2 k_0 \sqrt{\frac{R}{N (2+k_{0})(1+k_{0})^{R}}},
\label{ki-optimal}
\end{equation}
which gives the scaling $k_i^* \sim N^{-1/2}$ obtained from numerical
simulations [Fig.~\ref{scaling-optki}(a)].

It is also interesting to study the dependence of the optimal speed of
$I$ for different values of $U$'s diffusion rates $k_u^{\pm}$.
Numerical results are shown in Figure \ref{scaling-optki}(b), 
where we observe three different regimes in the behavior of
$k_i^*$ as a function of $k_u^+=k_u^-=k_u$:  
\begin{eqnarray*}
 k_i^* \simeq \left\{ \begin{array}{lll}
0.062\, k_u & \mbox{for} & 0 \le k_u \le 1, \\
0.062 & \mbox{for} & 1 \le k_u \le 2.062, \\
k_u-2 & \mbox{for} & 2.062 \le k_u \le 3.
\end{array}
\right.
\end{eqnarray*}
That is, $k_i^*$ is roughly constant for $1 \le k_u \le 2.062$ and grows
linearly with $k_u$ outside this range.  This result is quite
intriguing to us, as it is very simple but yet we cannot explain it intuitively.
Also, it seems to be quite hard to obtain an analytic estimation of
$k_i^*$ vs $k_u$.

\section{Two competing informed particles}
\label{twoinformed}

\begin{figure}
\centering 
\includegraphics[width=0.47\textwidth]{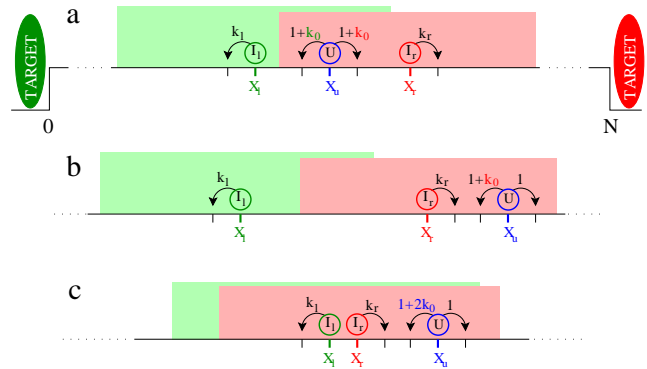}
\caption{(Color online).  Dynamics of two competing informed
  particles $I_r$ and $I_l$ traveling in opposite directions at
  speeds $k_r$ and $k_l$, respectively.  The uninformed particle $U$
  experiments a bias $k_0$ towards $I_r$ and/or $I_l$ when it is
  inside their respective attraction ranges, depicted by the red and
  green regions, 
  respectively.  $U$ symmetrically diffuses with rates $k_u^{\pm}=1$ outside
  that ranges, until it hits either right or left target.}  
\label{jumping-3}
\end{figure}

In Section \ref{oneinformed} we studied the simplest case of two
individuals searching for a target, where the leader
drags its uninformed partner towards a known target when they are
close enough. In this section we consider a more complex
situation, consisting of two informed individuals who try to recruit a
third uninformed partner. Each leader moves towards different targets located at the opposite
ends of the chain.  We call them right target $r$ (site $N)$ and
left target $l$ (site $0$).  The informed particles, denoted by
$I_r$ and $I_l$, start at the center site $c=N/2$ and move
ballistically to the right and left with rates
$k_r$ and $k_l$, respectively.  The uninformed particle $U$ also
starts at site $c$ and diffuses with rates that depend on its relative
position respect to $I_r$ and $I_l$, as in Section \ref{oneinformed}: $U$ has a
bias $k_0$ towards a given informed particle when it is within its 
interaction range, and diffuses symmetrically with rates
$k_u^+=k_u^-=1$ outside that range, as shown in Fig.~\ref{jumping-3}.
We can interpret this dynamics as two animals going to opposite
located food resources, and trying to convince an undecided
conspecific to follow them to their respective resources.

In order to make the analysis of this system as simple as possible we
consider two identical leaders that have the same interaction
range $R=10$ and attracting strength $k_0=0.2$, on a chain of length
$N=100$, and vary $k_r$ and $k_l$.  The question we want to explore
is: under which conditions one target (leader) is favored respect to the other
one? Or, how does the probability of reaching a given target depend on
the speeds $k_r$ and $k_l$?  

\begin{figure}
\centering \includegraphics[width=0.48\textwidth]{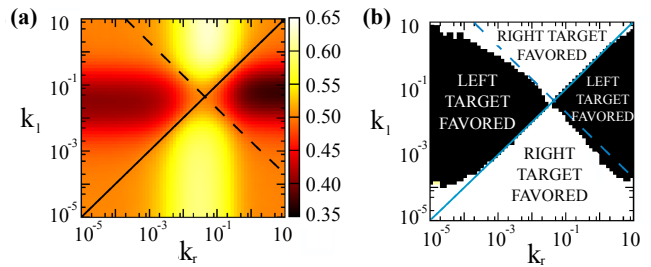}
\caption{(Color online) (a) Color-scale plot showing the probability
  $F^{\mbox{\tiny N}}$ that the uninformed particle arrives at the
  right target $N$ in presence of a right and a left moving particles, as a
  function of their speeds $k_r$ and $k_l$.  (b) Phase diagram showing
  the regions in the $k_l-k_r$ space where the uninformed particle most
  likely reaches the right (white: $F^{\mbox{\tiny N}}>1/2$) or left
  (black: $F^{\mbox{\tiny N}}<1/2$) target.}  
\label{both}
\end{figure}

Figure \ref{both}(a) shows in colored scale the probability
$F^{\mbox{\tiny N}}$ that particle $U$ reaches the right target $N$,
as a function of the speeds of the informed particles, $k_r$ and $k_l$.
Notice that the probability of getting the left target is simply
$1-F^{\mbox{\tiny N}}$.  The complementary plot in Fig. \ref{both}(b)
indicates the regions where a given target is 
favored.  White regions correspond to values of $k_r$ and $k_l$ where 
$F^{\mbox{\tiny N}} > 1/2$ (right target favored), while black regions
correspond to $F^{\mbox{\tiny N}} < 1/2$ (left target favored).   
As expected, along the $k_r=k_l$ line which separates
both regions is $F^{\mbox{\tiny N}}=1/2$ (solid line), corresponding
to the case of equally fast moving particles.  Perfectly identical
particles with the same initial conditions must have the same chances
to win. Interestingly, as we observe in Fig. \ref{both}, the
symmetric case $F^{\mbox{\tiny N}} = 1/2$ also happens for other
combinations of $k_r$ and $k_l$, indicated by the dashed line.  
An approximate expression for this crossover line can be obtained by
arguing that both informed particles would have the same likelihood to
guide $U$ as long as each particle has the same probability to guide $U$ to
its target independently, i.e., in the absence of the other informed
particle.  In other words, one can see the system of three particles
as two independent systems; one composed by $U$ and $I_r$, and the other
by $U$ and $I_l$.  Then, if $I_r$ drags $U$ in the first system with
the same probability as $I_l$ drags $U$ in the second system, then
$I_r$ and $I_l$ will drag $U$ with the same probability $1/2$ in the
combined 3-particle system.  This makes sense if we consider the case 
of one $I$ and one $U$ particles studied in section \ref{oneinformed}.  The  
relation between the FPP $F^{\mbox{\tiny N}}$ of particle $U$ and the 
speed $k_i$ of particle $I$ is plotted in Fig.~\ref{prob-ki}.  
Because of the non-monotonic shape of $F^{\mbox{\tiny N}}$, $U$ has the 
same hitting probability for two different speeds $k_i^1$ and $k_i^2$ of 
$I$, that is, $F^{\mbox{\tiny N}}(k_i^1)=F^{\mbox{\tiny N}}(k_i^2)$ [see
  Fig.~\ref{prob-ki}].  Therefore, we can arbitrarily 
identify these two speeds with the speeds $\tilde k_r$ and $\tilde
k_l$ of the right and left informed particles, and find the relation
which matches the FPPs to their targets.  Then, matching 
Eq.~(\ref{FN-low-ki}) for $F^{\mbox{\tiny N}}(\tilde k_r)$  with 
Eq.~(\ref{FN-high-ki}) for $F^{\mbox{\tiny N}}(\tilde k_l)$, in the low and high
speed limit, respectively, we obtain
\begin{eqnarray}
  \frac{(2+k_0)(1+k_0)^R \,\tilde k_r}{4 k_0} =  
  \frac{k_{0}R}{N \tilde k_l}.
\label{match}
\end{eqnarray}
Finally, pulling $\tilde k_r$ and $\tilde k_l$ to the left hand
side of Eq.~(\ref{match}), and using  
expression~(\ref{ki-optimal}) for the optimal value $k_i^*$, we arrive to 
\begin{eqnarray}
  \tilde k_r \, \tilde k_l = (k_i^*)^2.
\label{krkl}
\end{eqnarray}
Equation~(\ref{krkl}) gives an estimation of the nontrivial solution
corresponding to the dashed straight line in double logarithmic scale 
$\log \tilde k_l = 2 \log k_i^* -\log \tilde k_r$ of Fig.~\ref{both}, with
$k_i^*=0.062$ ($N=100, k_0=0.2, R=10$ and $k_u=1$). 

\begin{figure}
\centering  \includegraphics[width=0.45\textwidth]{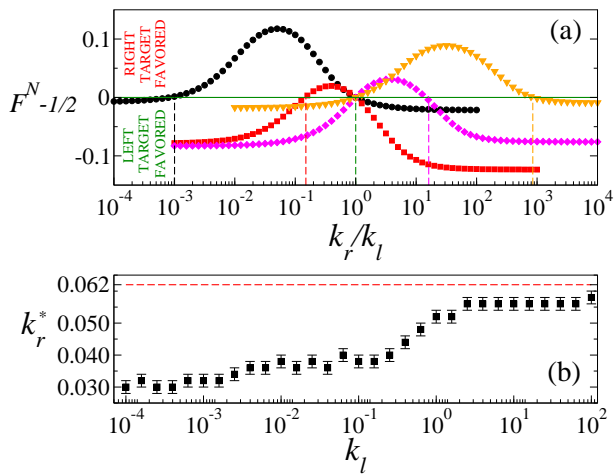}
\caption{(a) (Color online) Probability $F^{\mbox{\tiny N}}$ that the uninformed
  particle reaches the right target $N$ as a function of the ratio
  $k_{r}/k_{l}$ between the speeds of the right and left moving
  particles.  Each curve correspond to a fixed speed $k_l$ 
  [$k_l=1.0 (\bullet), k_l=10^{-1} (\blacksquare), k_l=10^{-2}
    (\blacklozenge)$ and $k_l=10^{-3} (\blacktriangledown)$]. Vertical dashed  
  lines denote the crossing points where $F^{\mbox{\tiny N}}=1/2$.  (b) Optimal
  speed of the right-moving particle $k_r^*$ that maximizes
  $F^{\mbox{\tiny N}}$ as a function of the speed $k_l$ of its competing
  partner.  The horizontal dashed line indicates the asymptotic value
  $k_r^*=0.062$ in the $k_l \to \infty$ limit.}
\label{comp2}
\end{figure}

In order to gain a deeper insight about the results reported in
Fig.~\ref{both}, we fix the speed of the left informed particle $k_l$
to a given value, and study how the FPPs behave as we vary the speed
of the right informed particle $k_r$.  This is equivalent  to
analyzing a horizontal cross section of the FPP landscape of
Fig.~\ref{both}.  Results for four different speeds $k_l$ are shown in
Fig.~\ref{comp2}(a), where we plot the probability of
reaching the right target $F^{\mbox{\tiny N}}$ as a function of the
ratio $k_r/k_l$.  We observe the following behavior as $k_r$ decreases from high
values.  For $k_r/k_l \gg 1$, the very fast moving particle $I_r$ has
a very short (almost negligible) interaction with $U$, and so we can
neglect the presence of $I_r$ and only consider the two-particle
system composed by $U$ interacting with $I_l$, like the one studied in
Section \ref{oneinformed}.  Therefore, the FPP to the left target
$F^{\mbox{\tiny 0}}(k_l)$ has the functional form of the
FPP plotted in Fig.~\ref{prob-ki}, where $k_l$ takes only the four
values $k_l=10^0, 10^{-1}, 10^{-2}$ and $10^{-3}$ of
Fig.~\ref{comp2}(a).  We make clear here that the $F^{\mbox{\tiny N}}$ 
vs $k_i$ curve of Fig.~\ref{prob-ki} corresponds to the FPP to the right
target when the informed particle $I$ moves right, but it must
be equivalent to the FPP to the left target when $I$ moves left.
Then, particle $U$ reaches the right target with the complementary
probability $F^{\mbox{\tiny N}}=1-F^{\mbox{\tiny 0}}(k_l)$, which agrees with the
asymptotic value of $F^{\mbox{\tiny N}}<1/2$ in the $k_r/k_l \gg 1$
limit [Fig.~\ref{comp2}(a)].  As $k_r$ decreases, $F^{\mbox{\tiny
    N}}$ increases, given that $I_r$ starts influencing the motion of
$U$, until it reaches a maximum.  Then, as $I_r$ moves even more
slowly $F^{\mbox{\tiny N}}$ decreases, reaching an asymptotic value in
the $k_r/k_l \ll 1$ limit corresponding to an almost static $I_r$.  We
also observe that $F^{\mbox{\tiny N}}$ takes the value $1/2$ in two
points, indicated by vertical dashed lines, one corresponding to $k_r
= k_l$ and the other to the nontrivial value approximated by
Eq.~(\ref{krkl}) (dashed line of Fig.~\ref{both}).  The theoretical
value $\tilde k_r = (k_i^*)^2/\tilde k_l$ agrees well with
numerical simulations only close to the $k_l=k_r=k_i^*$ point, as we can
see in Fig.~\ref{both}, where the solid and dashed lines cross.
However, discrepancies increase as we move away from $k_l=k_r=k_i^*$
  because the theoretical value of $k_i^*$ from Eq.~(\ref{ki-optimal})
  underestimates the numerical value, as Fig.~\ref{scaling-optki}(a) shows.

As we can see, the right-moving particle has to adapt its speed to the
speed of the left-moving competitor in order to have the greatest
chances to take the uninformed searcher to the right target.  A direct
consequence of this observation is the fact that the optimal speed
$k_r^*$ which maximizes $F^{\mbox{\tiny N}}$ depends on $k_l$.  This
is shown in Fig.~\ref{comp2}(b), where we see that $k_r^*$
increases with $k_l$.  In the high speed limit $k_l \to \infty$ of
the left particle, $I_l$ has no effect on $U$, thus $k_r^*$
asymptotically reaches the optimal value $k_i^*=0.062$ of the
two-particle system [see Fig.~\ref{comp2}(b)].  As $k_l$ decreases
from very high values, $I_l$ starts dragging $U$ to the left, so $I_r$
has to move slower to compensate this effect and maximize $F^{\mbox{\tiny N}}$,
monotonically reducing the value of $k_r^*$.

\section{Competing groups of informed particles}
\label{competegroups}

In the last section we studied the case in which two different
  leaders move in opposite directions.  But in a
more general scenario one can have two competing groups of leaders.
Our aim in this section is to explore what happens when the two groups
have a different size and persuading strength.
To that end we study a simple case consisting of two particles moving to the
left, one particle moving to the right, and one uninformed diffusing
particle $U$ looking for a target.  Naturally, the existence of
two leaders moving left increases the chances that the
uninformed particle reaches the left target.  However, as we shall
see, for some relations between the attraction strengths and speeds of
both groups, the minority may have the largest chances to win.

\begin{figure} 
\centering \includegraphics[width=0.48\textwidth]{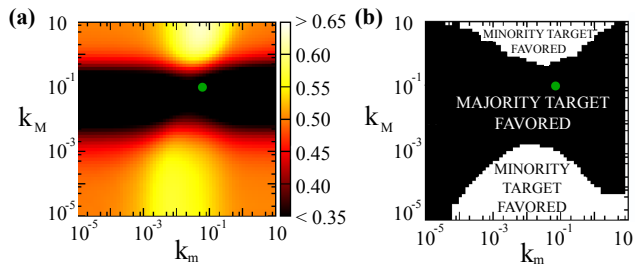}
\caption{(Color online) (a) Probability (color-scale)  
  that the uninformed particle arrives at the minority (right) target,
  under the influence of two informed particles moving to the left at
  speed $k_{\mbox{\tiny M}}$ and one to the right at speed $k_m$.  All three
  informed particles have the same internal strength $k_0=0.2$ and interaction
  range $R=10$.  (b) Regions in the
  $k_{\mbox{\tiny M}}-k_m$ speed space where the uninformed particle
  has the largest 
  chances to reach the minority (right) or majority (left) target.
  The largest region corresponds to the majority target favored.  The
  green dot indicates the optimal speed $k_{\mbox{\tiny M}}=k_m=0.062$.}
\label{tres}
\end{figure}

We start studying the case in which all three informed particles have
the same interaction strength $k_0=0.2$ and range $R=10$.  The
two-particle group (majority) moves left with speed 
$k_{\mbox{\tiny M}}$, while the 
one-particle group (minority) moves to the right with speed $k_m$.  Initially,
all four particles are at the center of 
the chain. The presence of a majority preferring the left target  
breaks the symmetry of the model studied in Section \ref{twoinformed}.
Indeed, Fig.~\ref{tres} shows that the majority (left) target has the largest
probability of being reached by $U$, for most combinations of
$k_{\mbox{\tiny M}}$ and 
$k_m$.  Only when the majority moves either too fast or too slow the
minority manages to take $U$ to the right target.  These two
limiting situations are slightly different.  In the 
$k_{\mbox{\tiny M}} \to 0$ limit, 
the left-moving majority remains static at the center and, therefore,
have no any bias effect on $U$.  But the right-moving minority
introduces a bias to the right, as we know from section
\ref{oneinformed}, giving an overall right bias that favors the
minority (right) target.  In the $k_{\mbox{\tiny M}} \to 
\infty$ limit the majority can be neglected,  
as it has a extremely short interaction with $U$, thus the minority
moving at finite speed has the largest chances to win.  Interestingly,
in the $10^{-3} \le k_{\mbox{\tiny M}} \le 2 \times 10^{-1}$ range,
the majority target is favored for all speeds $k_m$, that is, the
majority usually wins when moving at intermediate speeds, independent on the
minority's speed.  

\begin{figure}
\centering
\includegraphics[width=0.48\textwidth]{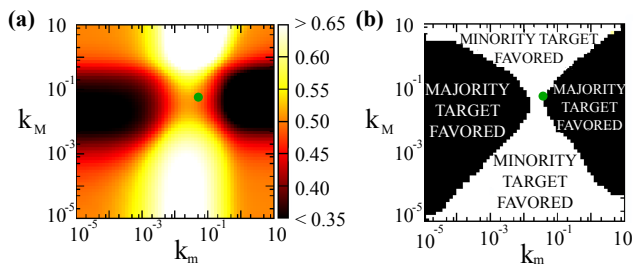}
\caption{ (Color online).  Same as in Fig.~\ref{tres} but for the case
  in which the strength $2k_0=0.4$ of the right-moving particle is
  twice the strength $k_0=0.2$ of each of the two left-moving
  particles.  The probability that the uninformed particle reaches the
  minority (right) target are largest for most combinations of speeds
  $k_{\mbox{\tiny M}}$ and $k_m$ (white region).}
\label{tresdist}
\end{figure}

Finally, it is interesting to study the case where the two groups have
different internal strengths.  In order to counterbalance the numerical
advantage of the majority group, we double the strength of the
particle in the minority, to the value $2 k_0=0.4$, but keep the
strength of both particles in the majority in the value $k_0=0.2$.
Then, the total majority strength $2 k_0$ acting on $U$ matches that of the
minority.  With these parameters, the two targets seem to be
equivalent, as it happens for the case of two identical informed particles
studied en Section \ref{twoinformed}.  However, as we observe in 
Fig.~\ref{tresdist}, the probability of arriving at the minority target
is largest for most $k_{\mbox{\tiny M}}-k_m$ combinations (white region), showing
that there is a preference for the minority target.  This is specially
evident around the $k_{\mbox{\tiny M}}=k_m=k_i^*=0.062$ region (green 
dot).  Indeed, when 
both the majority and the minority set their speeds equal and close to
the single-particle optimal speed $k_i^*$, the minority is 
more efficient in dragging $U$ to its target.  Seemingly, the fact that
the total strength $2 k_0$ of the majority is divided among two
particles causes a less effective dragging force on $U$ than that
caused by a single particle with strength $2k_0$.  An insight about
this result can be obtained with the following reasoning.  If both particles in
the majority would move together in perfect synchrony, then their
interaction ranges would always overlap along the path to the target,
and they would behave as a single particle with strength $2 k_0$.
However, stochasticity in the jumping process splits particles
apart, introducing two competing effects.  On the one hand, the
total interaction range composed by the range of the two left-moving
particles is larger than that of the right-moving particle, seemingly 
inducing an effective left drag over $U$.  But, on the other hand, the
strength of this total interaction range is $2k_0$ only where the ranges of
both left-moving particles overlap, and $k_0$ where they do not
overlap, which is only half of the strength of the right-moving
particle.  Apparently, this last mechanism induced by the difference in
strengths is stronger than the effect produced by enlarging the
interaction range, consequently breaking the symmetry in favor of the
right target.   

\section{Summary and conclusions}
\label{sec:conclusions}

In this article we have presented a minimalist approach to study how
the presence of a leader in a group of animals may influence the
movement of its conspecifics and incidentally 
their searching efficiency. The model has
two types of individuals: informed leaders that move ballistically at
speed $k_i$ towards the location of a known target, and uninformed
searchers that are attracted towards leaders when they are
close enough, but they freely diffuse with rate $k_u$ otherwise.  
To quantify the benefits of following the leader,  
we have focused on the probability $F^{\mbox{\tiny N}}$ that the uninformed
searcher finds a specific target $N$, and studied how that 
depends on $k_i$, $k_u$ and the internal attraction strength $k_0$.
   
We started analyzing the simplest case where there is only one
individual of each type.  We found that the first-passage
probability $F^{\mbox{\tiny N}}$ is non-monotonic and  
reaches its maximum at an intermediate value $k_i^*$ that is 
proportional to $k_u$.  Therefore, searching efficiency is maximized 
when the informed particle moves at intermediate speeds, as compared 
to the diffusion of the uninformed particle.  If it moves too
fast there is almost no interaction, and if it moves too slow the
interaction has almost no net effect on the final destination of the
uninformed particle, which is not able to take advantage of the
information exchange.  Moving not too fast but also not too slow seems to be the
optimal strategy to recruit undecided particles when they have a fixed
diffusion rate.

Then, we considered a more general situation in which two identical leaders
moving in opposite directions compete for an uninformed
individual.  It turns out that the 
particle that adapts its speed to a value closer to the optimal speed 
$k_i^*$ has the largest chances to win.  Surprisingly, a tie is
obtained for a nontrivial relationship between the speed of both
particles, besides the case where they move at the same speed.  When
a group of two leaders compete against a single particle group, the
largest group wins for most combinations of speeds, as its combined effective
persuading force is stronger.  However, this situation is reversed when
the social strength of the smaller group is doubled.  Therefore,
by following two different strategies a minority may be able to beat a
majority: either by increasing its internal strength or by adapting
its speed to an optimal value relative to the speed of the majority. 

It would be worthwhile to explore some extensions of the model that
include, for instance, many uninformed particles that interact not
only with the leaders but also between them.  Also, in a more
realistic scenario informed particles would not be identical but each would
have a different strength and range of interaction, and they may also
have some diffusion, instead of the pure ballistic motion considered in
the present work.

\section{Acknowledgments}
R.M-G. is supported by the JAEPredoc program of CSIC.
R.M-G. and C.L acknowledge support from MINECO
(Spain) and FEDER (EU) through Grants No. FIS2012-30634 (INTENSE@COSYP)
and CTM2012-39025-C02-01 (ESCOLA).

\appendix

\section{Calculation of the mean escape time from the attractive box $\tau_e$}
\label{appen-taue}

In this section we calculate the mean time $\tau_e$ that particle $U$ takes to
escape from the attracting box, when the box is static.  Given that
the position of the box is irrelevant in this case we assume, for
simplicity, that the box is centered at $n=0$.  Then $\tau_e$ is
equivalent to the mean exit time from an interval $[-R,R]$, starting 
from $n=0$ or, equivalently, the mean first-passage (MFPT) 
time to either site $n=-R-1$ or $n=R+1$.  Given that jumping rates are
symmetric around $n=0$, we can view the left half
interval $[-R-1,0]$ as a mirrored image of the right half $[0,R+1]$.
In this simpler scenario, we can consider the particle $U$ as confined
in a chain $[0,R+1]$ with a reflecting wall at $n=0$ and an absorbing site at
$n=R+1$.  Therefore, our problem is reduced to the
MFPT to site $R+1$, starting from
the reflecting boundary $0$, with jumping rates  
\begin{eqnarray}
k_u^+(n) &=& \left\{ \begin{array}{ll}
2 & \mbox{for $n=0$,} \\
1 & \mbox{for $1 \le n \le R$.}\end{array} \right. \nonumber \\
k_u^-(n) &=& 1+k_0 ~~~~ \mbox{for $1 \le n \le R$.} 
\label{kn+-reflect-1}
\end{eqnarray}
We note that the jumping rate $k_{0}^+=2$ at site $0$ is twice the
outgoing rate from $0$ in the complete chain.  The MFPT starting from
site $n$, $T(n)$, obeys the recursion equation \cite{gardiner}
\begin{equation}
k_u^+(n) \left[ T(n+1)-T_{n} \right] + k_u^-(n) \left[ T(n-1)-T(n) \right]=-1.
\label{rec-eq-1}
\end{equation}
The solution of Eq.~(\ref{rec-eq-1}) with reflecting and absorbing
boundary conditions $T(-1)=T(0)$ and $T(R+1)=0$ at $n=0$ and 
$n=R+1$, respectively, is given by \cite{gardiner}
\begin{eqnarray*}
T(n)&=& \sum_{y=n}^{R} \phi(y) \sum_{z=0}^{y} \frac{1}{k_u^+(z) \,
  \phi(z)},~~~\mbox{with} \\
\phi(y) &=& \prod_{z=1}^{y} \frac{k_u^-(z)}{k_u^+(z)}.
\end{eqnarray*}
Using rates in Eq.~(\ref{kn+-reflect-1}) gives $\phi(y) = r^y$ ($0 \le y \le
R$), where  
$r \equiv 1+k_0$ is the ratio between left and right rates.  Replacing
the expression for $\phi(y)$ and for the rates $k_u^+(n)$  from 
Eq.~(\ref{kn+-reflect-1}), the MFPT starting from site
$n=0$ can be expressed as
\begin{eqnarray} 
\tau_e \equiv T(0) = \sum_{y=0}^R r^y \sum_{z=0}^y \frac{1}{k_u^+(z) \,
  r^z} = \sum_{y=0}^R r^y \left[ \frac{1}{2} + \sum_{z=1}^y r^{-z} \right]. \nonumber \\
\label{taue-1}
\end{eqnarray}
To perform the sums in Eq.~(\ref{taue-1}) we make use of the equivalence 
$\sum_{k=0}^{M-1} \alpha^k = \frac{1-\alpha^M}{1-\alpha}$ for the sum
of a finite number of terms on a geometric series.  Finally, after doing
some algebra we arrive to 
\begin{eqnarray*} 
\tau_e=\frac{(r+1)(r^{R+1}-1)-2(r-1)(R+1)}{2(r-1)^2}, 
\end{eqnarray*}
or, in terms of the bias $k_0=r-1$,
\begin{eqnarray*} 
\tau_e = \frac{(2+k_0) \left[(1+k_0)^{R+1}-1 \right]-2 k_0 (R+1)}{2 k_0^2}, 
\end{eqnarray*}
which is the expression quoted in Eq.~(\ref{taue}).

\onecolumngrid

\section{Calculation of the mean exit time from the chain $T_c$}
\label{appen-T}

In this section we calculate the mean time that particle $U$ takes to
exit the chain, starting from the center site $c=N/2$, when the box is fixed and
centered at $c$.  Therefore, $U$ diffuses along the chain with
site-dependent jumping rates given by Eq.~(\ref{kn+-}). As we did
in Appendix \ref{appen-taue}, we take 
advantage of the system's symmetry around $c$ and map the chain $[0,N]$ with
absorbing boundaries at the extreme sites $n=0,N$, into the chain
$[c,N]$ with reflecting and absorbing boundaries at sites $n=c$
and $n=N$, respectively.  This mapping drastically reduces the
complexity of calculations.  Thus, the mean exit time corresponds to the
mean first-passage time (MFPT) to the absorbing boundary $N$, starting from
the reflecting boundary $c$, and jumping rates   
\begin{eqnarray}
k_u^+(n) &=& \left\{ \begin{array}{ll}
  2 & \mbox{for $n=c$,} \\
1 & \mbox{for $c+1 \le n \le N-1$.}\end{array} \right. \nonumber \\
k_u^-(n) &=& \left\{ \begin{array}{ll}
1+k_0 & \mbox{for $c+1 \le n \le c+R$,} \\
1 & \mbox{for $c+R+1 \le n \le N-1$.}\end{array} \right.
\label{kn+-reflect}
\end{eqnarray}
The rate $k_{u}^+(c)=2$ is twice the outgoing rate from $c$ in
the original chain.  The MFPT $T(n)$ starting from site $n$ obeys the
recursion formula 
\begin{eqnarray*}
k_u^+(n) \left[ T(n+1)-T_{n} \right] + k_u^-(n) \left[ T(n-1)-T(n) \right]=-1,
\label{rec-eq}
\end{eqnarray*}
whose solution with reflecting and absorbing boundary conditions 
$T_{c-1}=T_{c}$ and $T_{\mbox{\tiny N}}=0$, respectively, is given by
\cite{gardiner} 
\begin{eqnarray*}
T(n) = \sum_{y=n}^{N-1} \phi(y) \sum_{z=c}^{y} \frac{1}{k_u^+(z) \,
  \phi(z)}~~~~\mbox{with}~~~~ 
\phi(y) = \prod_{z=c+1}^{y} \frac{k_u^-(z)}{k_u^+(z)}.
\end{eqnarray*}
Using rates (\ref{kn+-reflect}) gives 
\begin{eqnarray}
\phi(y) = \left\{ \begin{array}{ll}
r^{y-c} & \mbox{for $c \le y \le c+R$,} \\
r^R & \mbox{for $c+R+1 \le y \le N-1$,}\end{array} \right. 
\label{phi}
\end{eqnarray}
where $r \equiv 1+k_0$.  Using rates $k_u^+(n)$ from 
Eqs.~(\ref{kn+-reflect}) and $\phi(y)$ from Eq.~(\ref{phi}), the MFPT 
starting from $c$ is 
\begin{eqnarray}
T_{c} = T(c)= \sum_{y=c}^{N-1} \phi(y) \sum_{z=c}^{y} \frac{1}{k_u^+(z)
  \,\phi(z)}   
=\sum_{y=c}^{c+R} r^{y-c} \left[ \frac{1}{2} +
  \sum_{z=c+1}^{y} r^{c-z} \right] + 
\sum_{y=c+R+1}^{N-1} r^R \left[ \frac{1}{2} + \sum_{c+1}^{c+R}
r^{c-z} + \sum_{z=c+R+1}^{y} r^{-R} \right], 
\end{eqnarray}
where we have split the sum over $z$ in site $c$ and intervals 
$[c+1,c+R]$ and $[c+R+1,y]$, and the sum over $y$ in intervals 
$[c,c+R]$ and $[c+R+1,N-1]$.  Performing the sums in brackets we obtain 
\begin{eqnarray*}
T_{c}&=&\sum_{y=c}^{c+R} r^{y-c} 
\left[ \frac{1}{2} + \frac{1-r^{c-y}}{r-1} \right]
+ \sum_{y=c+R+1}^{N-1} r^R \left[ \frac{1}{2} + \frac{1-r^{-R}}{r-1}  
  + r^{-R} \left( y-c-R \right)  \right]  \\
&=& \sum_{y=0}^{R} \left[ \frac{r^y}{2} + \frac{r^y-1}{r-1} \right] + 
\sum_{y=1}^{c-R-1} \left[ \frac{r^R}{2} + \frac{r^R-1}{r-1} + y \right]. 
\end{eqnarray*}
In the last equality, we have made a change of variables and
redistributed the terms.  Finally, performing the sums and replacing
back $r$ by $1+k_0$ we arrive to
\begin{equation}
 T_{c} = \frac{(2+k_0) \left[ (1+k_0)^{R+1} -1 \right] - 2 k_0
   (R+1)}{2 k_0^2}+
\frac{ (\tilde{N}-2) \left[ (2+k_0)(1+k_0)^R-2 \right] }
{4 k_0} + \frac{\tilde{N}(\tilde{N}-2)}{8},
\label{TL2}
\end{equation}
where $\tilde N \equiv N-2R$.  The first term in Eq.~(\ref{TL2})
exactly agrees with the calculated mean 
escape time from the box $\tau_e$ given by Eq.~(\ref{taue}).  Expressing
this first term as $\tau_e$ leads to the expression quoted in
Eq.~(\ref{T}).

\section{Estimation of the first-passage probability $F_{c+1}^{\mbox{\tiny N}}$}
\label{appen-E}

We now find an approximate expression for the FPP $F_{c+1}^{\mbox{\tiny N}}$ 
to target $N$ when the box jumps to site $n=c+1$ before particle $U$ exits the 
chain.  An exact calculation of  $F_{c+1}^{\mbox{\tiny N}}$ is hard to perform 
because this implies finding the occupation probability of $U$ along the chain for all
times.  It proves useful to divide the exit dynamics into two stages.
During the first stage $U$ diffuses along the chain with
jumping rates corresponding to the box centered at $c$, as
shown in Eq.~(\ref{kn+-}).  Then, the box jumps one step right and $U$
diffuses during a second stage with jumping rates given by 
\begin{eqnarray}
k_u^+(n) &=& \left\{ \begin{array}{ll}
1+k_0 & \mbox{for $c-R+1 \le n \le c$,} \\
1 & \mbox{otherwise,}\end{array} \right. \nonumber \\
k_u^-(n) &=& \left\{ \begin{array}{ll}
1+k_0 & \mbox{for $c+1 \le n \le c+R+1$,} \\
1 & \mbox{otherwise,}\end{array} \right.
\label{kn+-1}
\end{eqnarray}
until it hits one of the two targets.  Therefore, the probability to
hit target $N$ in a given realization of the dynamics depends on the
position of $U$ at the beginning of this second stage, that is, right
after the box moves.  Thus, $F_{c+1}^{\mbox{\tiny N}}$ can be estimated as 
\begin{equation}
F_{c+1}^{\mbox{\tiny N}} = \sum_{n=0}^N P(n) F_{c+1}^{\mbox{\tiny N}}(n),
\label{E-sum}
\end{equation}
where $P(n)$ is the probability that $U$ is at site $n$ when the box
moves to site $c+1$, and $F_{c+1}^{\mbox{\tiny N}}(n)$ is the FPP to
target $N$ starting from site $n$, when the box is centered at
$c+1$.  In Appendix \ref{appen-En} we show that 
\begin{eqnarray}
 F_{c+1}^{\mbox{\tiny N}}(n) \simeq \left\{ \begin{array}{ll}
 F_c^{\mbox{\tiny N}}(n)+ \frac{1}{\tilde{N}} & 
\mbox{for $c-R+1 \le n \le c+R$,} \\
 F_c^{\mbox{\tiny N}}(n)  & \mbox{otherwise,} \end{array} \right. 
\label{En}
\end{eqnarray}
where $ F_c^{\mbox{\tiny N}}(n)$ is the FPP to target $N$ starting
from site $n$, with the box centered at $c$.   These FPPs obey the 
boundary conditions $F_{c}^{\mbox{\tiny N}}(0)=F_{c+1}^{\mbox{\tiny N}}(0)=0$ 
and $F_{c}^{\mbox{\tiny N}}(N)=F_{c+1}^{\mbox{\tiny N}}(N)=1$.
The following two properties prove useful in performing the sum of
Eq.~(\ref{E-sum}):   
\begin{eqnarray}
\label{PL2}
&&P(c-n)=P(c+n)~~~\mbox{and}~~~ \\
&& F_c^{\mbox{\tiny N}}(c-n)+ F_c^{\mbox{\tiny N}}(c+n)=1
~~~\mbox{for $0 \le n \le c$},
\label{EL2}
\end{eqnarray}
which reflect the symmetry of the system and initial conditions
around $c$.  That is, given that $U$ diffuses during the first stage
under a symmetric landscape of rates, the occupation probability $P(n)$ must
be symmetric around $c$ [Eq.~(\ref{PL2})].  In addition,
Eq.~(\ref{EL2}) reflects 
the fact that the exiting probabilities 
through $0$ and $N$, starting from the same distance to those
borders must be equal, as we explicitly show in Appendix \ref{appen-En}.
From relations (\ref{PL2}) and (\ref{EL2}) one finds 
\begin{equation}
\sum_{n=0}^N P(n)  F_c^{\mbox{\tiny N}}(n)=1/2.
\label{sum-P}
\end{equation}
Then, using expressions (\ref{En}) for $F_{c+1}^{\mbox{\tiny N}}(n)$ in the 
sum of Eq.~(\ref{E-sum}), and the relation (\ref{sum-P}), we obtain 
\begin{eqnarray}
 F_{c+1}^{\mbox{\tiny N}} = \sum_{n=0}^N P(n) F_c^{\mbox{\tiny $N$}}(n) 
 + \frac{1}{\tilde{N}} \sum_{n=c-R+1}^{c+R} P(n) 
 \simeq \frac{1}{2} + \frac{P_{\mbox \tiny box}}{\tilde{N}},
\label{E-sum-1}
\end{eqnarray}
where we have defined 
$P_{\mbox \tiny box} \equiv \sum_{n=c-R}^{c+R} P(n)$, as the
probability that $U$ is inside the box when the box jumps.  
An approximate expression for $P_{\mbox \tiny box}$ can be obtained
by noting that the likelihood that $U$ is inside the box, i e., 
in the interval $[c-R,c+R]$, when it has not reached any target yet,
should be proportional to 
the rate at which $U$ enters the box $8/\tilde{N}$, as compared to the
rate at which $U$ leaves the box $1/\tau_e$ (see Appendix
\ref{appen-Pbox}).  Therefore, we arrive to 
\begin{equation}
P_{\mbox \tiny box} \simeq \frac{8 \tau_e}{8\tau_e + \tilde{N}}.
\label{Pbox}
\end{equation}
Finally, plugging Eq.~(\ref{Pbox}) for $P_{\mbox \tiny box}$ into Eq.~(\ref{E-sum-1}) 
for $F_{c+1}^{\mbox{\tiny N}}$ we arrive to the expression quoted in Eq.~(\ref{Fc1}) of 
the main text.

\section{Estimation of the first-passage probability $F_{c+1}^{\mbox{\tiny N}}(n)$}
\label{appen-En}

We calculate in this section the probability 
$F_{c+1}^{\mbox{\tiny N}}(n)$ that particle $U$ hits target $N$
starting from site $n$, when the box is centered at site $c+1$.
In principle, we expect this probability to be very similar to the
hitting probability $ F_c^{\mbox{\tiny N}}(n)$ starting from site
$n$ and with the box centered at $c$, instead of $c+1$.  In fact,
as we shall see, these two probabilities 
only differ in a small ``perturbation'' of order $1/\tilde{N}$.  We first
illustrate how to calculate $ F_c^{\mbox{\tiny N}}(n)$, and then
apply the same technique to calculate $ F_{c+1}^{\mbox{\tiny N}}(n)$. 

Given that FFPs are unequivocally determined by the right and left jumping 
probabilities at different sites $p_n^+$ and $p_n^-$, respectively, it
turns out more convenient to work in discrete time.  That is, we see
particle $U$ as making either a right or a left step in a time
interval $\Delta t=1/K$, where $K=2+k_0$ ($K=2$) is the total jumping
rate when $U$ is inside (outside) the box.  From Eq.~(\ref{kn+-}), jumping
probabilities can be written as 
\begin{eqnarray*}
p_n^+ &=& \left\{ \begin{array}{lll}
p & \mbox{for $c-R \le n \le c-1$,} \\
q & \mbox{for $c+1 \le n \le c+R$,} \\
\frac{1}{2} & \mbox{otherwise,}\end{array} \right. \nonumber \\
p_n^- &=& \left\{ \begin{array}{lll}
q & \mbox{for $c-R \le n \le c-1$,} \\
p & \mbox{for $c+1 \le n \le c+R$,} \\
\frac{1}{2} & \mbox{otherwise,}\end{array} \right.
\label{p+-}
\end{eqnarray*}
where $p \equiv (1+k_0)/(2+k_0)$ and $q \equiv 1/(2+k_0)$.  Note
that $p+q=1$.  $ F_c^{\mbox{\tiny N}}(n)$ obeys the following
recursion equations
\begin{eqnarray}
F(n)&=&\frac{1}{2} F(n-1) + \frac{1}{2} F(n+1)~~~~~\mbox{for~~~~ $1 \le n
  \le c-R-1$,} \nonumber \\
F(n)&=&q F(n-1) + p F(n+1)~~~~~~ \mbox{for~~~~ $c-R \le n \le c-1$,}
\nonumber \\
F(n)&=&\frac{1}{2} F(c-1) + \frac{1}{2} F(c+1)~~~~~~
\mbox{for~~~~ $n=c$,} \nonumber \\
F(n)&=&p F(n-1) + q F(n+1)~~~~~~ \mbox{for~~~~ $c+1 \le n \le
  c+R$,} \nonumber \\
F(n)&=&\frac{1}{2} F(n-1) + \frac{1}{2} F(n+1)~~~~~ \mbox{for~~~~
  $c+R+1 \le n \le N-1$,} 
\label{En-L2}
\end{eqnarray}
where we have dropped indices $c$ and $N$ to simplify notation.  The
solution to the system of equations (\ref{En-L2}), subject to absorbing boundary
conditions at site $0$ and $N$, $F(0)=0$ and $F(N)=1$, respectively, is given by
\begin{eqnarray}
 F_c^{\mbox{\tiny N}}(n) = \left\{ \begin{array}{lllll}
A(r-1) n & 
\mbox{for~~~~ $0 \le n \le c-R$,} \\
\frac{A}{2} \left[ (r-1)\tilde{N} + 2 \left(1-r^{c-R-n} \right) \right] &
\mbox{for~~~~ $c-R \le n \le c$,} \\ 
\frac{1}{2} & 
\mbox{for~~~~ $n=c$,} \\
\frac{A}{2} \left[ (r-1) \tilde{N} + 2 \left(1-2r^{-R} \right) +
  2r^{n-c-R} \right] &
\mbox{for~~~~ $c \le n \le c+R$,} \\
A \left[ (r-1)(n-2R) + 2\left(1- r^{-R}\right) \right] & 
\mbox{for~~~~ $c+R \le n \le N$,} 
\end{array} \right.
\label{En-L2-2}
\end{eqnarray}
where $r \equiv p/q = 1+k_0$ and 
$A \equiv \left[ (r-1)\tilde{N}+2 \left(1-r^{-R} \right) \right]^{-1}$. 
We have also placed back indices $c$ and $N$.  We can check that
$ F_c^{\mbox{\tiny N}}(c-n)+ F_c^{\mbox{\tiny N}}(c+n)=1$, with
$0 \le n \le c$, the symmetry property expressed in Eq.~(\ref{EL2}) of 
appendix \ref{appen-E}.

To calculate $ F_{c+1}^{\mbox{\tiny N}}(n)$ we follow the same method
as for $ F_c^{\mbox{\tiny N}}(n)$, but with jumping probabilities
corresponding to the box centered at $c+1$
\begin{eqnarray*}
p_n^+ &=& \left\{ \begin{array}{lll}
p & \mbox{for $c-R+1 \le n \le c$,} \\
q & \mbox{for $c+2 \le n \le c+R+1$,} \\
\frac{1}{2} & \mbox{otherwise.}\end{array} \right. \nonumber \\
p_n^- &=& \left\{ \begin{array}{lll}
q & \mbox{for $c-R+1 \le n \le c$,} \\
p & \mbox{for $c+2 \le n \le c+R+1$,} \\
\frac{1}{2} & \mbox{otherwise.}\end{array} \right.
\label{p+-1}
\end{eqnarray*}
We obtain
\begin{eqnarray*}
 F_{c+1}^{\mbox{\tiny N}}(n) = \left\{ \begin{array}{lllll}
A(r-1) n & 
\mbox{for~~~~ $0 \le n \le c-R+1$,} \\
\frac{A}{2} \left[ (r-1)\tilde{N} + 2r \left(1-r^{c-R-n} \right) \right] &
\mbox{for~~~~ $c-R+1 \le n \le c+1$,} \\ 
\frac{A}{2} \left[ (r-1)\tilde{N}+2r \left(1-r^{-R-1} \right) \right] & 
\mbox{for~~~~ $n=c+1$,} \\
\frac{A}{2} \left[ (r-1) \tilde{N} + 2r \left(1-2r^{-R-1} \right) +
  2r^{n-c-R-1} \right] &
\mbox{for~~~~ $c+1 \le n \le c+R+1$,} \\
A \left[ (r-1)(n-2R) + 2 \left(1- r^{-R}\right) \right] & 
\mbox{for~~~~ $c+R+1 \le n \le N$.} 
\end{array} \right.
\label{En-L2-3}
\end{eqnarray*}
$ F_{c+1}^{\mbox{\tiny N}}(n)$ can be written in terms of 
$ F_c^{\mbox{\tiny N}}(n)$ from Eq.~(\ref{En-L2-2}) as  
\begin{eqnarray}
 F_{c+1}^{\mbox{\tiny N}}(n) =  F_c^{\mbox{\tiny N}}(n) + \left\{ \begin{array}{lllll}
0 & 
\mbox{for~~~~ $0 \le n \le c-R$,} \\
A (r-1) \left(1-r^{c-R-n} \right) &
\mbox{for~~~~ $c-R \le n \le c$,} \\ 
A (r-1) \left(1-r^{-R} \right) & 
\mbox{for~~~~ $n=c$,} \\
A (r-1) \left( 1-r^{n-c-R-1} \right) &
\mbox{for~~~~ $c+1 \le n \le c+R+1$,} \\
0 & 
\mbox{for~~~~ $c+R+1 \le n \le N$.} 
\end{array} \right.
\label{En-L2-4}
\end{eqnarray}  
Finally, expanding Eq.~(\ref{En-L2-4}) to first order in
$1/r=1/(1+k_0) < 1$, $ F_{c+1}^{\mbox{\tiny N}}(n)$ can be reduced to the
simple approximate expression
\begin{eqnarray*}
 F_{c+1}^{\mbox{\tiny N}}(n) \simeq  F_c^{\mbox{\tiny N}}(n) +
\left\{ \begin{array}{lll} 
0 & 
\mbox{for~~~~ $0 \le n \le c-R$,} \\
\frac{1}{\tilde{N}} &
\mbox{for~~~~ $c-R+1 \le n \le c+R$,} \\
0 & 
\mbox{for~~~~ $c+R+1 \le n \le N$,} 
\end{array} \right.
\label{En-L2-5}
\end{eqnarray*}  
quoted in Eq.~(\ref{En}) of appendix \ref{appen-E}.

\twocolumngrid

\section{Estimation of the occupation probability inside the box $P_{box}$}
\label{appen-Pbox}

We calculate here an approximate expression for the probability
$P_{box}$ that particle $U$ is inside the box, i.e., located at a site $n$
in the range $c-R \le n \le c+R$, when the box jumps one step
right.  We shall see that $P_{box}$ can be estimated as the ratio
between the rates associated to $U$ entering and leaving the box. 

Within a very simplified coarse-grained picture of the system, we
consider that if particle $U$ did not exit the chain, it can be in
only two possible occupation states, either inside the box 
(state $1$) or outside the box (state $0$).  That is, states $1$ and $0$ 
correspond to $U$ being at sites $|n-c| \le R$ and $|n-c| > R$,
respectively.  Occupation probabilities $P_0(t)$ and $P_1(t)$ of
states $0$ and $1$ at time $t$ evolve following these master equations
\begin{eqnarray}
\frac{\partial P_0(t)}{\partial t} = k_{10} P_1(t) - k_{01} P_0(t),
\nonumber \\
\frac{\partial P_1(t)}{\partial t} = k_{01} P_0(t) - k_{10} P_1(t),
\label{dpdt}
\end{eqnarray}
where $k_{10}$ ($k_{01}$) is the transition rate from inside (outside)
to outside (inside) the box.  The total occupation probability is
normalized to one [$P_0(t)+P_1(t)=1$], because we restrict to the case
where the particle is still inside the chain.  If we run many
realizations of the dynamics, this means that at a given time $t$ we
only consider those realizations in which $U$ did not exit the
chain, and associate $P_1(t)$ with the fraction of those
realizations where $U$ is inside the box.  The solution to
Eqs.~(\ref{dpdt}) with initial condition $P_i(t)=\delta_{i,1}$ ($U$
inside the box) is 
\begin{eqnarray}
P_0(t) = \frac{k_{10} \left[1-e^{-(k_{01}+k_{10})t}\right]}{k_{01}+k_{10}},
\nonumber \\
P_1(t) = \frac{k_{01} + k_{10} e^{-(k_{01}+k_{10})t}}{k_{01}+k_{10}}.
\label{P1-t}
\end{eqnarray}
To estimate $P_{box}$ we assume that, after some time,
the occupation probability at site $n$, $P(n)$, used to derive 
$ F_{c+1}^{\mbox{\tiny N}}$, reaches a stationary value.  Therefore, we
associate $P_{box}$ with the stationary value of $P_1$ from
Eq.~(\ref{P1-t}) in the long time limit, that is,
\begin{eqnarray}
P_{box} \simeq P_1^s = \frac{k_{01}}{k_{01}+k_{10}}.
\label{P1st}
\end{eqnarray}
The outgoing rate $k_{10}$ can be estimated as the inverse of the mean escape
time from the box 
\begin{eqnarray}
k_{10} \simeq \frac{1}{\tau_e}, 
\label{k10}
\end{eqnarray}
with $\tau_e$ given by Eq.~(\ref{taue}).  Now, to estimate the incoming
rate $k_{01}$ we focus on the situation where $U$ just leaves the box
through the left side, jumping from site $n=c-R$ to site $n=c-R-1$.
Once in site $n=c-R-1$, $U$ performs a symmetric random walk in the interval
$[0,c-R]$ until it either returns back to site $n=c-R$ or hits the
absorbing site $n=0$.  If we denote by $p_r$ the returning probability
and by $T_r$ the mean time to exit the interval, then the returning rate
can be approximated as $p_r/T_r$.  And, given that $U$ can escape
through either side of the box, the incoming rate is twice the
returning rate, and so 
\begin{eqnarray*}
k_{01} \simeq 2p_r/T_r.
\end{eqnarray*}
To calculate $p_r$ and $T_r$ we consider $U$ jumping with equal rates 
$k_u^+(n)=k_u^-(n)=1$ in the interval $[0,c-R]$.  In
this context, $p_r$ is the FPP to site $c-R$ and $T_r$ is the mean
first-passage time (MFPT) to sites $0$ or $c-R$, starting from site $c-R-1$.  

The FPP starting from site $n$, $F_n$, obeys the recursion equation 
\begin{eqnarray*}
F(n) = \frac{1}{2} F(n-1) + \frac{1}{2} F(n+1),
\label{rec-eq-E}
\end{eqnarray*}
whose solution with absorbing boundary conditions $F(0)=0$ and $F(c-R)=1$ is  
\begin{eqnarray*}
F(n) = \frac{n}{c-R}.
\label{En-1}
\end{eqnarray*}
Therefore, the returning probability is
\begin{eqnarray}
p_r = F(c-R-1) = \frac{c-R-1}{c-R}.
\label{pr}
\end{eqnarray}
Also, the MFPT starting from site $n$, $T(n)$,
obeys a similar recursion equation
\begin{eqnarray*}
T(n+1) + T(n-1) - T(n) = -1,
\label{rec-eq-T}
\end{eqnarray*}
whose solution with absorbing boundary conditions $T(0)=T(c-R)=0$ is 
\begin{eqnarray*}
T(n) = \frac{n(c-R-n)}{2},
\end{eqnarray*}
and, therefore, 
\begin{eqnarray}
T_r = T(c-R-1) = \frac{c-R-1}{2}.
\label{Tr}
\end{eqnarray}
Then, combining Eqs.~(\ref{pr}) and (\ref{Tr}), the incoming rate
can be approximated as 
\begin{eqnarray}
k_{01} \simeq \frac{2 p_r}{T_r} \simeq \frac{8}{\tilde{N}},
\label{k01}
\end{eqnarray}
where we have replaced back $c-R$ by $N/2-R=\tilde N/2$.
Finally, plugging expressions (\ref{k10}) and (\ref{k01}) for
$k_{10}$ and $k_{01}$, respectively, into Eq.~(\ref{P1st}) we obtain
the expression 
\begin{eqnarray*}
P_{box} \simeq \frac{8 \tau_e}{8 \tau_e+\tilde{N}},
\end{eqnarray*}
quoted in Eq.~(\ref{Pbox}) of appendix \ref{appen-E}.


\end{document}